%% file: neurips_2026.tex
\documentclass{article}

 \usepackage[preprint]{neurips_2026}

\usepackage[utf8]{inputenc} %
\usepackage[T1]{fontenc}    %
\usepackage{hyperref}       %
\usepackage{url}            %
\usepackage{booktabs}       %
\usepackage{amsfonts}       %
\usepackage{nicefrac}       %
\usepackage{microtype}      %
\usepackage{xcolor}         %

\usepackage{amsmath}
\usepackage{siunitx}
\usepackage{wrapfig}
\usepackage{graphicx}
\usepackage{subfigure}
\usepackage{multirow, makecell}
\usepackage{enumitem}
\usepackage{xspace}
\usepackage{acronym}
\usepackage{amsmath}
\usepackage[capitalize,noabbrev]{cleveref}
\usepackage{tikz}
\usetikzlibrary{calc}
\usepackage{pifont}
\usepackage{newunicodechar}
\newunicodechar{✓}{\ding{51}}
\newunicodechar{✗}{\ding{55}}

\title{Exploring Token-Space Manipulation\\ in Latent Audio Tokenizers}

\author{%
  Francesco Paissan$^{1,2}$,
  Luca Della Libera$^{2,3}$,
  Mirco Ravanelli$^{2,3}$,
  Cem Subakan$^{1,2}$\thanks{Correspondence to \texttt{francesco.paissan@mila.quebec}.} \\
  $^1$Mila -- Qu\'ebec AI Institute, $^2$Universit\'e Laval, $^3$Concordia University
}

\begin{document}

\maketitle

\acrodef{rvq}[RVQ]{residual vector quantization}
\acrodef{bsq}[BSQ]{Binary Spherical Quantization}
\acrodef{ola}[OLA]{overlap-add}
\acrodef{mos}[MOS]{mean opinion score}
\acrodef{dnsmos}[DNSMOS]{Deep Noise Suppression mean opinion score}
\acrodef{utmos}[UTMOS]{UTokyo-SaruLab mean opinion score}
\acrodef{dwer}[dWER]{differential word error rate}
\acrodef{asr}[ASR]{automatic speech recognition}
\acrodef{snr}[SNR]{signal-to-noise ratio}

\begin{abstract}
  Neural audio codecs provide compact discrete representations for speech generation and manipulation. However, most codecs organize tokens as frame-level sequences, making it difficult to study or intervene on global factors of variation. In this work, we propose the Latent Audio Tokenizer for Token-space Editing (LATTE) that appends a fixed set of learnable latent tokens to the audio feature sequence and retains only these tokens for quantization and decoding. This design produces a compact, non-temporally aligned bottleneck in which each token can aggregate global information across the full utterance. We show that the resulting tokenizer preserves competitive reconstruction quality in low-bitrate speech coding settings while enabling simple token-space interventions. In particular, we find that swapping selected latent token positions between utterances can modify global attributes, such as speaker identity and background noise, and we evaluate these interventions on voice conversion and denoising tasks. Our results suggest that compact latent audio tokenizers can support controllable audio manipulation without supervision in task-specific editing models. Audio samples are available at \href{https://fpaissan.github.io/latte-website/}{this link}. %
\end{abstract}

\section{Introduction}
\input{sections/introduction}

\input{sections/related}
\input{sections/methods}
\input{sections/importance}
\input{sections/experiments}

\input{sections/results}

\section{Conclusions}
\label{sec:conclusion}

We introduced LATTE, a TiTok-style latent audio tokenizer that compresses speech into a compact set of non-temporally aligned discrete slots. Despite this global bottleneck, LATTE preserves competitive low-bitrate resynthesis quality and exposes a token interface suitable for analysis and intervention. Slot-importance profiles show that latent positions are not exchangeable: different factors concentrate on different subsets of slots, with noise profiles showing the strongest stability and speaker-related attributes remaining partially entangled. Using these profiles for token replacement enables simple zero-shot denoising and speaker-transfer interventions, outperforming random and least-important slot controls at matched edit budgets.

Overall, our results suggest that compact latent audio tokenizers can serve not only as efficient compression interfaces, but also as probes of global factor structure in audio representations. The current evidence supports partial specialization rather than hard disentanglement, and the token-swap results should be viewed as controlled interventions rather than full editing systems. Scaling latent-slot tokenization to more diverse audio domains and developing objectives that encourage cleaner factor separation are natural directions for future work.

\newpage

\begingroup
\bibliographystyle{plainnat}
\bibliography{bibliography}
\endgroup
\clearpage
\appendix

\section{Limitations and Broader Impacts}
\label{sec:limitations}

\paragraph{Limitations.}
A limitation of the present LATTE model is the scale and diversity of its training data. Although the architecture builds on a strong pretrained WavLM front-end, the LATTE-specific components inherited from the FocalCodec pipeline are trained on only a few hundred hours of clean English speech. This is substantially lower than that of recent competing tokenizers such as WavTokenizer, StableCodec, and Mimi, which are trained on much larger and more diverse datasets. This data gap may affect not only reconstruction quality and robustness, but also the reliability with which individual latent slots specialize to interpretable factors such as speaker identity, background noise, or channel conditions. Future work should therefore scale LATTE training to multilingual speech, noisy and reverberant conditions, mixtures, and broader audio domains to test whether the observed slot structure becomes more stable and general.

\paragraph{Broader Impacts.}
Controllable speech tokenization can support useful applications such as speech
restoration, privacy-preserving factor analysis, and more efficient generative
audio modeling. The same capability can also be misused for voice conversion or
speaker-identity manipulation. We therefore frame the token-swap experiments as
analysis tools rather than deployment-ready conversion systems, and any release
of checkpoints should be accompanied by clear usage restrictions, provenance
tracking, and detection or watermarking mechanisms where appropriate.

\section{Additional Slot-Importance Diagnostics}
\label{app:slot_importance_extra}

\begin{figure}[t]
    \centering
    \includegraphics[width=0.85\linewidth]{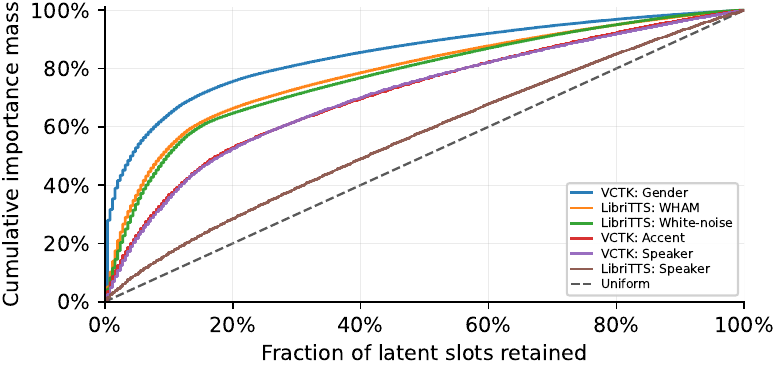}
    \caption{
        Cumulative importance mass after sorting slots by descending importance.
        Curves above the uniform baseline indicate that a factor is concentrated
        in a subset of slots. The absence of near-step-function behavior supports
        partial specialization rather than hard slot-factor disentanglement.
    }
    \label{fig:cumulative_importance}
\end{figure}

\paragraph{Cumulative Concentration.}
Figure~\ref{fig:cumulative_importance} complements the entropy and Gini
statistics in Table~\ref{tab:importance_structure} by showing how quickly each
importance profile accumulates mass when slots are sorted from most to least
important. All curves lie above the uniform baseline, confirming that factor
information is not spread evenly across latent positions. At the same time, none
of the curves approaches a step function: even the most concentrated attributes
require multiple slots to account for most of the importance mass. This supports
the interpretation that \textsc{LATTE} develops partially specialized latent
positions, rather than a hard one-slot-per-factor decomposition.

\begin{figure*}[t]
    \centering
    \subfigure[LibriTTS: Noise]{
        \includegraphics[width=0.46\textwidth]{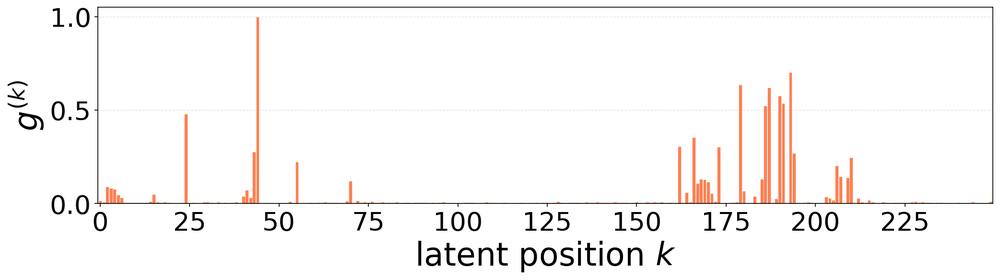}
    }\hfill
    \subfigure[LibriTTS: Speaker]{
        \includegraphics[width=0.46\textwidth]{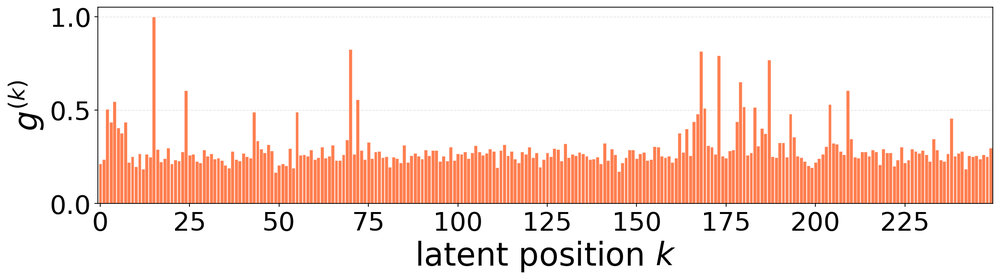}
    }\\[0.75em]

    \subfigure[VCTK: Accent]{
        \includegraphics[width=0.46\textwidth]{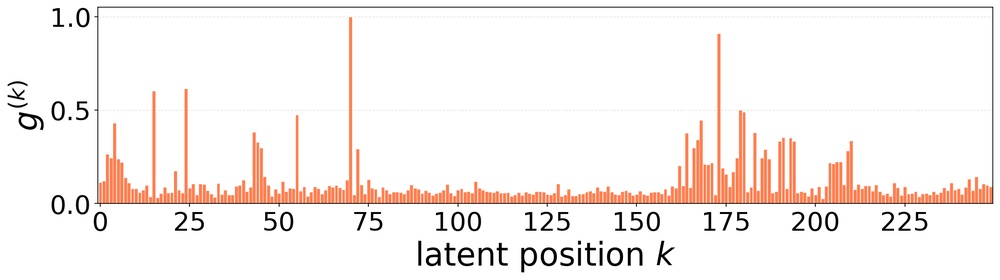}
    }\hfill
    \subfigure[VCTK: Speaker]{
        \includegraphics[width=0.46\textwidth]{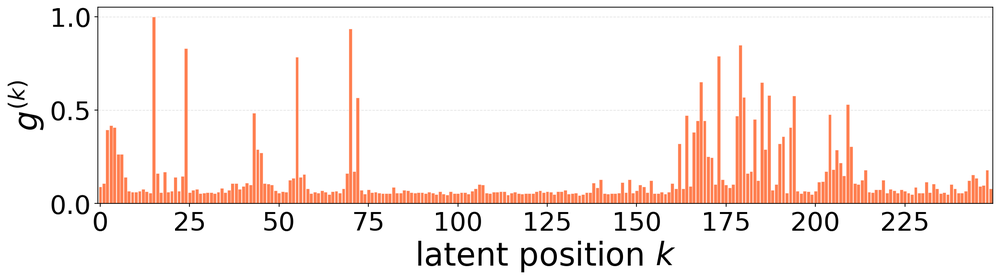}
    }\\[0.75em]

    \subfigure[VCTK: Gender]{
        \includegraphics[width=0.46\textwidth]{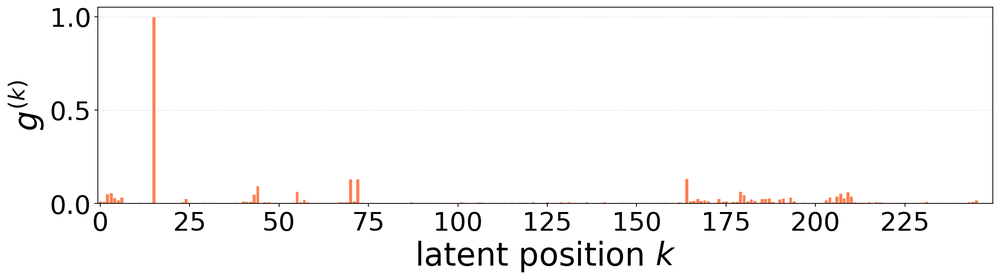}
    }
    \caption{
        Per-factor importance scores for \textsc{LATTE}. Higher values indicate
        latent positions whose mean codes vary more strongly across the
        corresponding factor partition.
    }
    \label{fig:importance_scores_lat}
\end{figure*}

\paragraph{\textsc{LATTE} Slot Profiles.}
Figure~\ref{fig:importance_scores_lat} provides the unnormalized per-slot
importance profiles underlying the aggregate results in
Figure~\ref{fig:cumulative_importance} and Table~\ref{tab:importance_structure}.
The profiles show that high-importance slots are sparse but not isolated:
several factors activate small groups of latent positions, and related factors
often share part of their support. Noise profiles on LibriTTS exhibit recurring
high-importance regions, consistent with the strong white-noise/WHAM! agreement
reported in the main text. Speaker, accent, and gender profiles show more overlap,
especially on VCTK, supporting our conclusion that speaker-related attributes are
structured but partially entangled.

\begin{figure*}[t]
    \centering
    \subfigure[LibriTTS: Noise]{
        \includegraphics[width=0.46\textwidth]{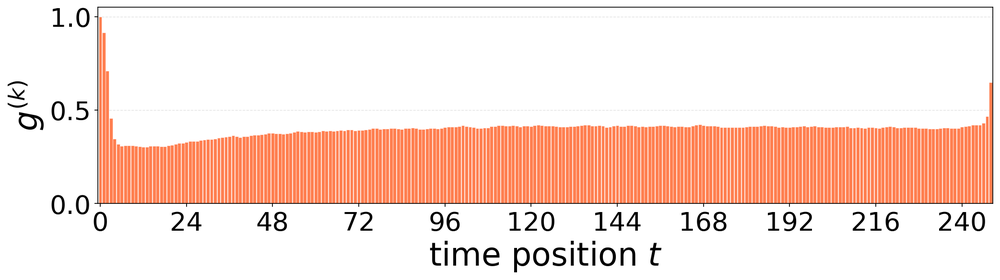}
    }\hfill
    \subfigure[LibriTTS: Speaker]{
        \includegraphics[width=0.46\textwidth]{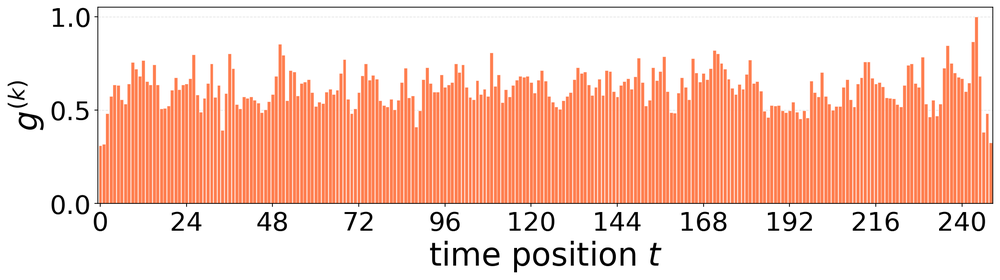}
    }\\[0.75em]

    \subfigure[VCTK: Accent]{
        \includegraphics[width=0.46\textwidth]{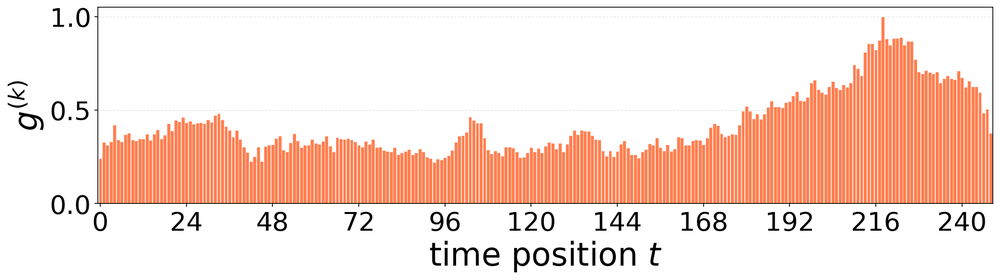}
    }\hfill
    \subfigure[VCTK: Speaker]{
        \includegraphics[width=0.46\textwidth]{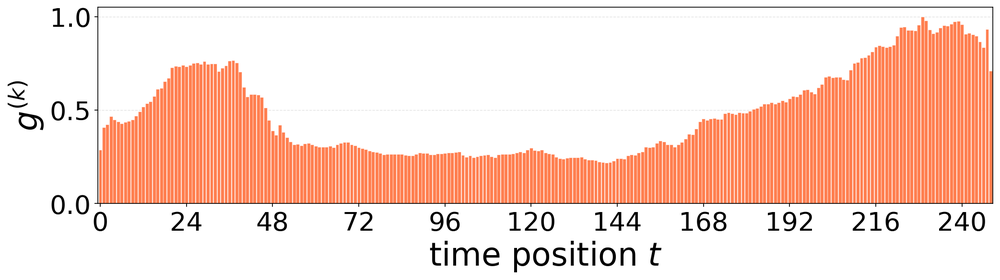}
    }\\[0.75em]

    \subfigure[VCTK: Gender]{
        \includegraphics[width=0.46\textwidth]{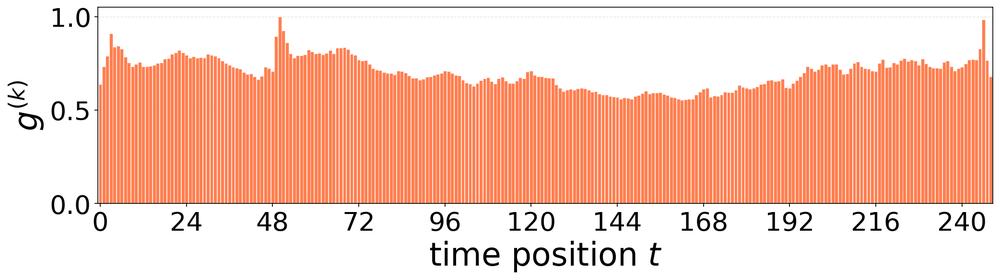}
    }
    \caption{
        Per-factor importance scores for FocalCodec tokenization, computed with
        the same partition-based scoring procedure.
    }
    \label{fig:importance_scores_focalcodec}
\end{figure*}

\paragraph{Comparison with Frame-Level Tokenization.}
Figure~\ref{fig:importance_scores_focalcodec} applies the same scoring procedure
to the original frame-level FocalCodec tokenization. Compared with
Figure~\ref{fig:importance_scores_lat}, these profiles are less naturally
interpretable as slot-level factors, since token positions are tied to temporal
frames rather than fixed latent roles. This contrast is important: in
\textsc{LATTE}, a high-importance index refers to the same latent slot across
utterances, making it meaningful to rank and replace positions for token-space
intervention. In the frame-level representation, high scores instead reflect
where factor-dependent variation appears over time, which is less directly
suited to global attribute swapping.

\section{Evaluation Checkpoints and Metric Details}
\label{app:metric_checkpoints}

This appendix summarizes the external models, checkpoints, and preprocessing used for evaluation. Unless otherwise stated, all audio is converted to mono and resampled to 16 kHz before computing metrics. Our implementation is based on the \texttt{audiocodecs} repository\footnote{\url{https://github.com/lucadellalib/audiocodecs/}}.

\paragraph{UTMOS.}
For LibriSpeech test-clean, we report UTMOS as a non-reference perceptual speech-quality metric. We use the \texttt{utmos22\_strong} model loaded via \texttt{torch.hub} from \texttt{tarepan/SpeechMOS:v1.2.0}. UTMOS is computed directly on the hypothesis waveform at 16 kHz.

\paragraph{DNSMOS.}
For VoiceBank and Libri1Mix, we report DNSMOS. We use the implementation bundled with the evaluation code, which loads the checkpoint and computes a P.808-style MOS estimate from Mel features. The final score is averaged over analysis windows. We do not use DNSMOS P.835. In our resynthesis evaluation, DNSMOS is used only for VoiceBank and Libri1Mix; LibriSpeech uses UTMOS.

\paragraph{dWER.}
We report dWER as an audio-only measure of linguistic preservation. Unless a reference transcript is explicitly provided, both the reference waveform and the hypothesis waveform are transcribed with \texttt{faster-whisper}. We use \texttt{WhisperModel} with the default model hub identifier \texttt{small}, corresponding to Whisper small, and the tokenizer \texttt{openai/whisper-small}. The ASR device and compute type are configurable; the default compute type is \texttt{int8}. Both ASR outputs are normalized with \texttt{WhisperTokenizer.normalize}, tokenized by whitespace, and compared using \texttt{SpeechBrain}'s \texttt{ErrorRateStats}. Thus, dWER is the WER between the two ASR transcripts, not the difference between two WER values.

\paragraph{Speaker Similarity.}
We report speaker similarity (Sim) using a pretrained speaker-verification model. We use the WavLM speaker-verification backend, \texttt{microsoft/wavlm-base-sv}.

\paragraph{Baseline and Model Checkpoints.}
For the original FocalCodec baselines, we use the public \texttt{lucadellalib/focalcodec} hub models. Unless otherwise stated, non-FocalCodec baseline numbers in the resynthesis tables follow the common FocalCodec evaluation protocol. For all the other baselines, refer to the \texttt{audiocodecs} repository. %

\input{sections/appendix_editing_protocols}

\input{sections/appendix_ola}

\input{sections/appendix_titok_hyperparams}

\end{document}

%% file: sections/introduction.tex
Neural audio codecs have become the dominant paradigm in speech representation learning, generation, and manipulation~\citep{zeghidour2021soundstream,defossez2023encodec,kumar2023dac} with applications to the multimodal domain~\citep{dubey2024llama3herdmodels,jiang2024mixtralexperts,comanici2025gemini25,singh2025openaigpt5card,deepseekai2025deepseekv3}.
By compressing waveforms into sequences of discrete tokens, they provide a 
universal language that connects acoustic signal processing with modern language modeling pipelines~\citep{borsos2023audiolm, kreuk2023audiogen, wang2023valle, nguyen2024spiritlminterleavedspokenwritten, defossez2024moshi, kyutai2025hibiki, zeghidour2025delayed}.
The dominant approach consists of generating \emph{frame-level} token streams: one or more codes per fixed-duration frame, organized into long, locally aligned sequences via \ac{rvq}~\cite{zeghidour2021soundstream}.
This structure is well-suited to waveform fidelity and left-to-right generation, but it limits how easily one can study or intervene on \emph{global} factors of variation, such as speaker identity, background noise, or speaking style.
Because each token represents a narrow temporal window, isolating or manipulating a global attribute requires identifying \emph{which} frames carry it, an inherently diffuse and challenging problem.

\begin{figure}[t]
  \centering
  \includegraphics[width=.95\columnwidth]{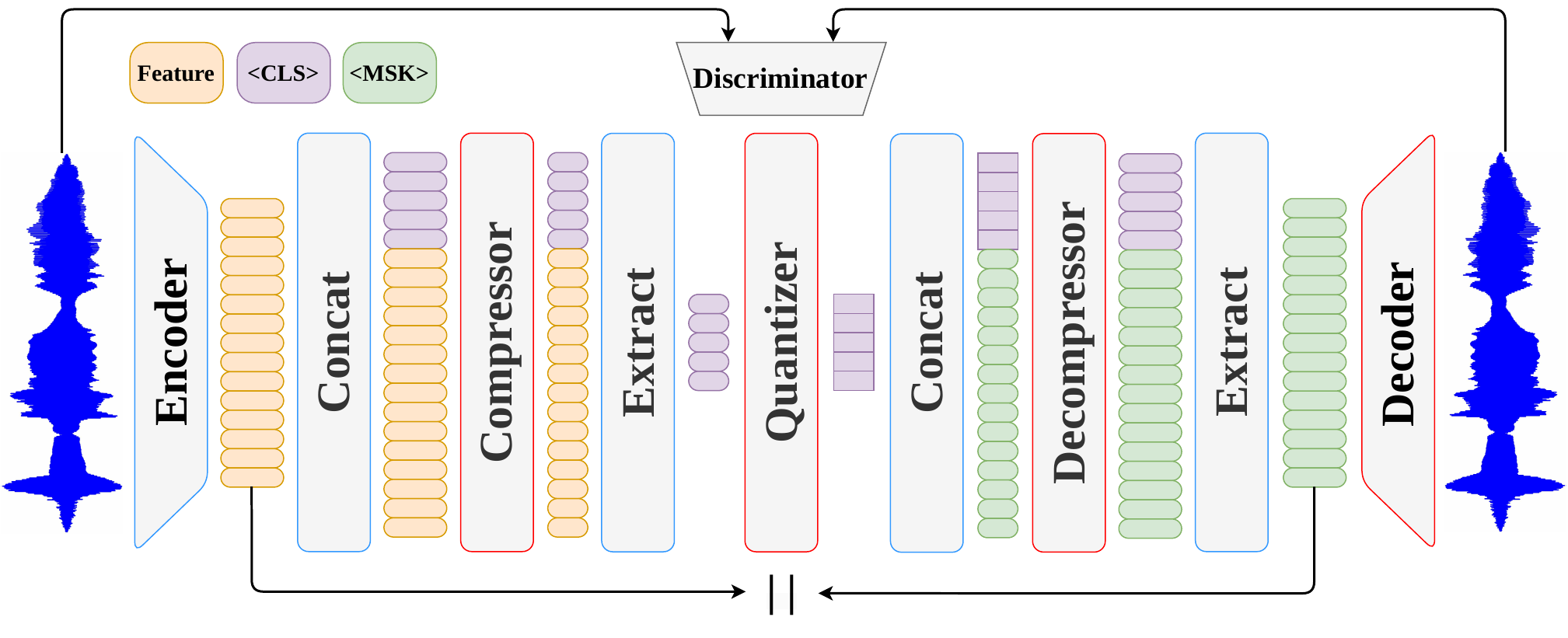}
  \vspace{-10pt}
  \caption{\textbf{LATTE turns frame-level codec features into compact latent slots for analysis and editing.} A frozen FocalCodec front-end maps speech to acoustic-semantic features; a learned TiTok-style bottleneck keeps only quantized latent slots; importance scores identify slots associated with global factors, which can then be swapped between utterances for zero-shot speaker and noise edits. Rounded edges denote continuous representations, squared edges denote discrete representations; blue borders indicate frozen modules, while red borders indicate that the model is learned. \texttt{Feature} represent WavLM-layer 6 features. \texttt{<CLS>} represent the learnable latent tokens, appended to the sequence of latent audios. \texttt{<MSK>} represents the learnable mask tokens used as initial point for reconstruction.}
  \label{fig:overview}
  \vspace{-10pt}
\end{figure}

A complementary line of work takes a different structural approach: rather than representing audio as a long sequence of frames, it compresses the entire utterance into a set of learned latent tokens that serve as a bottleneck before reconstruction.
TiTok~\citep{yu2024an} introduced this idea for images, replacing patch-level tokens with a fixed number of latent queries that aggregate global information before discrete quantization.
ALMTokenizer~\citep{yang2025almtokenizer} adapted a similar strategy to audio, combining masked autoencoding and autoregressive objectives to produce a compact token sequence suited to language modeling.
These architectures suggest a natural hypothesis: if the model must reconstruct the full signal from a small set of non-temporally aligned slots, individual slots may specialize to encode distinct global factors, possibly leading to a structured latent space that is more transparent and more amenable to targeted intervention.

We investigate this hypothesis directly.
Concretely, we adapt FocalCodec~\citep{dellalibera2025focalcodec}, a low-bitrate, single-codebook speech codec based on WavLM~\citep{chen2022wavlm} features and focal modulation~\citep{yang2022focalnets,dellaliber2024focal}, into a TiTok-style latent audio tokenizer. While the same pipeline can be applied to different architectures, we chose FocalCodec for its effective and efficient representation of both acoustic and semantic content.
We keep the FocalCodec encoder and decoder frozen, and insert a learned compressor--quantizer--decompressor chain between them.
The compressor appends $L$ learned latent queries to the codec feature sequence and runs a shared self-attention encoder; only the query positions are retained, discretized with \ac{bsq}~\citep{zhao2024bsq}, and passed to a symmetric decoder that reconstructs the full codec feature sequence.
Because no temporal alignment is imposed, each latent slot is free to pool information from the entire utterance, possibly inducing \emph{interpretable specialization} in the learned slots, independently of codec-specific inductive biases.

Our contributions are a follow:
\begin{itemize}[leftmargin=1.5em, topsep=2pt, itemsep=1pt]
    \item We introduce LATTE, a TiTok-style latent audio tokenizer that compresses an entire utterance into a set of global discrete tokens, enabling compact and partially controllable speech representations. Despite compressing a full utterance through a compact, low-bitrate discrete bottleneck, the proposed tokenizer preserves competitive reconstruction quality on standard low-bitrate speech-coding benchmarks, matching or approaching frame-level baselines.
    \item We perform a Jaccard-relevance analysis of the individual latent positions, showing that they are \emph{not} uniformly involved across factor types: some slots concentrate mass relevant to noise or speaker identity, while others are more diffuse. This pattern is qualitatively different from standard frame-level tokens.
    \item We use slot-importance scoring to analyze and intervene in global factors, enabling zero-shot speaker and noise-token swaps. We find that swapping selected latent token positions between two utterances produces targeted shifts in global attributes: transferring noise-relevant slots changes the background acoustic environment, and transferring speaker-relevant slots partially transfers speaker identity, without supervised editing models or test-time optimization.
\end{itemize}

Taken together, these results show that the new tokenization strategy used in LATTE achieves performance competitive with state-of-the-art low-bitrate baselines, while also offering strong controllability and interpretability.
Audio samples are available at this link\footnote{\url{https://fpaissan.github.io/latte-website/}.}.

%% file: sections/related.tex
\section{Related Work}

\paragraph{Speech Codecs.}
Most speech tokenizers discretize audio at a fixed temporal resolution~\citep{mousavi2025discrete}. Early neural codecs emphasized acoustic fidelity at moderate bitrates, using
convolutional encoder--decoder models and \ac{rvq} to reconstruct waveforms
from frame-level code streams~\citep{zeghidour2021soundstream,defossez2023encodec,kumar2023dac}.
These acoustic codecs preserve fine detail, but their long, locally aligned
token sequences make global attributes such as speaker identity or background
condition diffuse across time. Semantic tokenizers make the opposite trade-off: they quantize self-supervised speech features, such as HuBERT or WavLM, to
produce compact linguistic units for language modeling~\citep{hsu2021hubert,chen2022wavlm,mousavi2026dasb}.
This improves semantic abstraction but discards prosody, speaker cues, and
fine acoustic detail that must later be recovered by a decoder.
Recent hybrid codecs aim to balance semantic usefulness and acoustic
reconstruction. SpeechTokenizer~\citep{zhang2024speechtokenizer} and Mimi~\citep{defossez2024moshi} distill semantic information into codec tokens;
SemantiCodec~\citep{liu2024semanticodec} combines semantic tokens with a
generative acoustic decoder; and WavTokenizer~\citep{ji2024wavtokenizer},
BigCodec~\citep{xin2024bigcodec}, Stable Codec~\citep{parker2024scaling}, and
FocalCodec~\citep{dellalibera2025focalcodec,dellalibera2025focalcodecstream}
explore low-bitrate, single-codebook, or streamable designs with stronger token
structure. Our work is orthogonal to this fixed-rate progression: rather than
improving a frame-aligned token stream, we ask whether a compact set of
non-temporally aligned latent slots can expose and manipulate global factors.

\paragraph{Compact Latent Bottlenecks.}
Highly compressed tokenizers replace long patch- or frame-level sequences with
a small set of learned tokens that aggregate information before quantization.
TiTok~\citep{yu2024an} introduced this idea for images by processing patches
jointly with learned latent queries and keeping only the query outputs for
vector quantization. ALMTokenizer~\citep{yang2025almtokenizer} adapts related
ideas to audio, using learnable queries, semantic-prior initialization, masked
autoencoding, and autoregressive objectives to produce compact sequences for
audio language modeling. We do not directly compare with ALMTokenizer because
public code and checkpoints are unavailable at the time of submission. Instead,
we isolate the representation question: whether this learned-token bottleneck
induces factor-dependent structure that can be probed and edited.

\paragraph{Latent Space Controllability.}
Discrete latent spaces can support manipulation when individual token subsets
have predictable functional roles \citep{Beyer2025HighlyCT}. Image tokenizers such as TiTok demonstrate
that compact learned tokens can retain enough global information for generation
and reconstruction~\citep{yu2024an}; our question is whether analogous audio
tokens expose factor structure that can be intervened on directly. We therefore use an audio-domain causal probe: estimate which latent slots are associated
with a global factor, swap those slots across utterances, and measure whether
the decoded waveform moves along the intended attribute.

%% file: sections/methods.tex
\section{Latent Audio Tokenizer} \label{sec:method}

We use FocalCodec~\citep{dellalibera2025focalcodec} as a frozen speech-codec substrate and
replace its original frame-level bottleneck with a learned latent-token
bottleneck. This lets us test whether non-temporally aligned discrete slots
develop an interpretable specialization while keeping the codec feature space and
waveform decoder fixed. An overview of the pipeline is presented in \cref{fig:overview}.

Let $\mathbf{x}\in\mathbb{R}^{S}$ denote a speech segment sampled at
$f_s=\SI{16}{\kilo\hertz}$, with $S=f_s\tau$ for chunk duration $\tau$
($\tau=\SI{5}{\second}$ by default). FocalCodec maps the
waveform to WavLM layer-6 features~\citep{chen2022wavlm}, yielding a
representation space rich in acoustic and semantic content. We denote the
frozen front-end as
\begin{equation}
    \mathcal{E}_{\mathrm{FC}}:\mathcal{X}\rightarrow\mathbb{R}^{T\times H},
    \qquad
    \mathbf{F}=\mathcal{E}_{\mathrm{FC}}(\mathbf{x}),
\end{equation}
where $H=1024$ and $T=50\tau$. We insert a compressor--quantizer--decompressor
chain between the frozen encoder and decoder:
\begin{equation}
    \mathbf{x}
    \xrightarrow{\ \mathcal{E}_{\mathrm{FC}}\ }
    \mathbf{F}
    \xrightarrow{\ g_\theta\ }
    \mathbf{Z}
    \xrightarrow{\ q_\phi\ }
    \mathbf{C},\mathbf{k}
    \xrightarrow{\ h_\psi\ }
    \hat{\mathbf{F}}
    \xrightarrow{\ \mathcal{D}_{\mathrm{FC}}\ }
    \hat{\mathbf{x}},
\end{equation}
where $g_\theta$ is a learned compressor, $q_\phi$ a discrete quantizer that
returns quantized vectors and their integer indices, and $h_\psi$ a learned
decompressor; $\mathcal{E}_{\mathrm{FC}}$ and
$\mathcal{D}_{\mathrm{FC}}$ remain frozen throughout training.

\paragraph{Latent Slot Compression.}
The compressor summarizes the codec feature sequence into $L$ learned latent
slots~\citep{yu2024an,yang2025almtokenizer}. We set $L = r\tau$, where $r$ is the target
token rate; the default $r=\SI{50}{\hertz}$ gives $L=250$ tokens per chunk.

We introduce $L$ learned queries $\mathbf{Q}\in\mathbb{R}^{L\times H}$,
concatenate them with the positionally-embedded codec features, and process
the joint sequence through a FocalCodec-style encoder:
\begin{equation}
\mathbf{Y}
=
\left[
    \mathbf{F}+\mathbf{P}^{\mathrm{feat}}_{1:T}
    \ ;\
    \mathbf{Q}+\mathbf{P}^{\mathrm{lat}}_{1:L}
\right],
\qquad
    \mathbf{Z}
    =
    g_\theta(\mathbf{Y})_{T+1:T+L}
    \in\mathbb{R}^{L\times d},
\end{equation}
where $\mathbf{P}^{\mathrm{feat}}$ and $\mathbf{P}^{\mathrm{lat}}$ are
learned positional embeddings. Only the $L$ latent positions are retained, so each slot can pool information from the full feature sequence while reconstruction must flow exclusively through these $L$ vectors.

\paragraph{Binary Spherical Quantization.}
Each latent vector $\mathbf{z}_\ell\in\mathbb{R}^{d}$ is quantized with
\ac{bsq}~\citep{zhao2024bsq}, yielding a discrete
code $k_\ell\in\{0,\dots,2^d-1\}$ and a quantized vector
$\mathbf{c}_\ell\in\mathbb{R}^{d}$. Collecting over all slots:
\begin{equation}
    \mathbf{C}\in\mathbb{R}^{L\times d},
    \qquad
    \mathbf{k}\in\{0,\dots,2^d-1\}^{L}.
\end{equation}
We use $d=13$ (vocabulary size $2^{13}=8192$); at rate $r$ this gives a
bitrate of $rd$ bits/s---\SI{650}{\bit\per\second} at default settings.
\Ac{bsq} also contributes an auxiliary regularizer $\mathcal{L}_{\mathrm{BSQ}}$
that promotes balanced codebook usage.

\paragraph{Feature Reconstruction.}
The decompressor maps the quantized sequence back to the original frame rate.
Its input concatenates $T$ learned reconstruction slots with the
positionally-embedded quantized codes:
\begin{equation}
\mathbf{U}
=
\left[
    \mathbf{m}\otimes\mathbf{1}_{T}
    +
    \mathbf{P}^{\mathrm{mask}}_{1:T}
    \ ;\
    \mathbf{C}
    +
    \mathbf{P}^{\mathrm{code}}_{1:L}
\right],
\end{equation}
where $\mathbf{m}\in\mathbb{R}^{d}$ is a learned mask embedding and
$\mathbf{P}^{\mathrm{mask}}$, $\mathbf{P}^{\mathrm{code}}$ are learned
positional embeddings. Reconstructed codec features are read from the first
$T$ output positions:
\begin{equation}
    \hat{\mathbf{F}}
    =
    h_\psi(\mathbf{U})_{1:T}
    \in\mathbb{R}^{T\times H}.
\end{equation}

\paragraph{Training Objective.}
The tokenizer minimizes reconstruction error in the frozen codec feature space:
\begin{equation}
    \mathcal{L}(\theta,\phi,\psi)
    =
    \left\|
        \hat{\mathbf{F}} - \mathbf{F}
    \right\|_2^2
    +
    \lambda \mathcal{L}_{\mathrm{BSQ}},
\end{equation}
where the squared-error term is averaged over non-padded frames and $\lambda$
weights the quantization regularizer. We use no waveform-domain loss, and the
FocalCodec encoder and decoder are not updated.

%% file: sections/importance.tex
\section{Token Importance Scoring} \label{sec:importance}

TiTok-style bottlenecks create a fixed set of latent positions, but they do not
specify what each position should encode. We therefore use slot importance as a
post-hoc probe of factor structure~\citep{Beyer2025HighlyCT}. For a factor of variation such as speaker
identity, accent, or noise level, we assign a scalar relevance score $g_\ell$ to
every latent index $\ell\in\{1,\dots,L\}$.
Given a dataset annotated with a factor label, utterances are partitioned into
$J$ groups. For each partition $j$ we accumulate the sample-mean code vector at
every token position:
\begin{equation}
    \boldsymbol{\mu}_{j,\ell}
    =
    \frac{1}{N_j}\sum_{n\in\mathcal{P}_j}\mathbf{c}_{n,\ell}
    \;\in\mathbb{R}^{d},
\end{equation}
where $\mathcal{P}_j$ is the index set of partition $j$ and $N_j=|\mathcal{P}_j|$.
We then form the matrix $\mathbf{M}_\ell\in\mathbb{R}^{J\times d}$ whose $j$-th
row is $\boldsymbol{\mu}_{j,\ell}$, and let
$\mathbf{X}_\ell = \mathbf{M}_\ell - \mathbf{1}_J\bar{\boldsymbol{\mu}}_\ell^{\!\top}$
be its row-centred form, where
$\bar{\boldsymbol{\mu}}_\ell = \tfrac{1}{J}\sum_j\boldsymbol{\mu}_{j,\ell}$.
The importance score is then
\begin{equation}
    g_\ell
    =
    \frac{\sigma_{\max}(\mathbf{X}_\ell)^{2}}{J-1},
    \label{eq:importance}
\end{equation}
where $\sigma_{\max}(\mathbf{X}_\ell)$ is the largest singular value of
$\mathbf{X}_\ell$. This equals the leading eigenvalue of the sample
between-partition covariance of the mean codes at position $\ell$, and measures
how much the mean code at that slot spreads across factor groups in all $d$
dimensions simultaneously. A high score indicates that position $\ell$ carries
factor-specific information; a low score indicates factor-invariance.

Scores are computed separately for each factor of interest, yielding one
importance vector $\mathbf{g}=(g_1,\dots,g_L)\in\mathbb{R}^{L}$ per factor. To characterize
how concentrated factor information is across slots, we treat $\mathbf{g}$ as
an unnormalized distribution and compute entropy, Gini coefficient, and
Jaccard overlap between importance vectors from different factor partitions.
These metrics reveal the degree to which individual slots specialize to
distinct global attributes rather than spreading information uniformly.

The importance vectors also define a mechanism for zero-shot
attribute transfer. Given a source utterance with quantized codes
$\mathbf{C}^{(s)}\in\mathbb{R}^{L\times d}$ and a target utterance with codes
$\mathbf{C}^{(t)}$, we first rank the token positions by decreasing
factor importance. Let $\pi$ be a permutation such that
$g_{\pi_1}\geq g_{\pi_2}\geq \dots \geq g_{\pi_L}$. We then select the smallest
set of top-ranked positions whose cumulative importance reaches a prescribed
fraction $\gamma$:
\begin{equation}\label{eq:swap1}
    m_\gamma
    =
    \min \left\{
        m :
        \sum_{i=1}^{m} \bar{g}_{\pi_i} \geq \gamma
    \right\},
    \qquad
    \mathcal{S}_\gamma
    =
    \{\pi_1,\dots,\pi_{m_\gamma}\},
\end{equation}
where $\bar{g}_\ell = g_\ell / \sum_{j=1}^{L} g_j$ denotes the normalized
importance score. We construct the swapped code matrix as
\begin{equation}\label{eq:slot2}
    \tilde{\mathbf{C}}^{(s)}_\ell
    =
    \begin{cases}
        \mathbf{C}^{(t)}_\ell & \ell \in \mathcal{S}_\gamma, \\
        \mathbf{C}^{(s)}_\ell & \text{otherwise.}
    \end{cases}
\end{equation}
The edited code matrix is then mapped back to codec features by $h_\psi$ and
decoded by $\mathcal{D}_{\mathrm{FC}}$ to produce the edited waveform. The same
operation can equivalently be viewed as replacing the corresponding discrete
indices $\mathbf{k}$ before codebook lookup. %

%% file: sections/experiments.tex
\section{Experiments}
\label{sec:experiments}

We evaluate \textsc{LATTE} along three axes. First, we measure resynthesis quality to verify that its compact latent representation preserves intelligibility, speaker information, and perceptual quality under standard codec benchmarks. Second, we analyze the learned latent slots by computing slot-wise importance scores for several annotated global factors: speaker identity, speaker gender, accent, and background noise level. Third, we evaluate whether these importance scores can be used to select slots for predictable token-space edits.

Our experiments are designed to answer the following questions:
(i) Does \textsc{LATTE} remain competitive with existing neural codec tokenizers in resynthesis quality?
(ii) Do different latent slots exhibit non-uniform associations with interpretable audio factors?
(iii) Can these slot-factor associations be used to perform targeted edits while preserving non-target properties?

\subsection{Datasets and Tasks}

All audio is resampled to 16 kHz. For resynthesis, we follow the evaluation protocol of FocalCodec~\citep{dellalibera2025focalcodec} and report results on LibriSpeech test-clean~\citep{panayotov2015librispeech}, VoiceBank~\citep{valentinibotinhao2016voicebank}, and Libri1Mix~\citep{cosentino2020librimix}. These benchmarks cover clean speech, noisy speech, and speech mixtures, allowing us to evaluate reconstruction quality across increasingly challenging acoustic conditions. 

We compare \textsc{LATTE} against a broad set of neural codec and speech-tokenizer baselines, including EnCodec~\citep{defossez2023encodec}, DAC~\citep{kumar2023dac}, WavLM6-KM~\citep{chen2022wavlm}, SpeechTokenizer~\citep{zhang2024speechtokenizer}, SemantiCodec~\citep{liu2024semanticodec}, Mimi~\citep{defossez2024moshi}, WavTokenizer~\citep{ji2024wavtokenizer}, BigCodec~\citep{xin2024bigcodec}, Stable Codec~\citep{parker2024scaling}, and FocalCodec at 50 Hz~\citep{dellalibera2025focalcodec}. Unless otherwise stated, baseline numbers follow the common FocalCodec evaluation pipeline; checkpoint and metric details are listed in \cref{app:metric_checkpoints}.

For slot-importance analysis, we construct factor partitions from dataset metadata or controlled corruptions. For speaker identity, each speaker is treated as a separate class, and we compute partitions on both VCTK~\citep{veaux2017cstr} and LibriTTS~\citep{zen2019libritts} to evaluate cross-dataset stability. For speaker gender, we use LibriTTS and VCTK metadata. For accent, we use VCTK labels. For noise-level analysis, we corrupt LibriTTS utterances with additive white noise and WHAM! noise at 0, 5, 10, 20, and 40 dB \ac{snr}, testing whether the same latent slots are associated with degradation level across noise distributions.

For token-space editing, we evaluate speaker- and noise-intervention methods. Speaker editing uses a parallel VCTK voice-conversion setting: the target is a clean, different-speaker utterance with the same sentence as the source. Denoising uses non-self-clean targets selected by a deterministic cyclic shift over the corresponding evaluation list, so the target is clean but generally not text- or speaker-matched. Full pairing details are provided in \cref{app:editing_protocols}.

We evaluate two asymmetric encoder--decoder configurations, denoted \textsc{LATTE Base} and \textsc{LATTE Large}. In our notation, the first label indicates encoder scale and the second indicates decoder scale: BL corresponds to a Base encoder with a Large decoder, while \textsc{LATTE Large} uses a Large encoder with an XL decoder. Both variants share the same tokenization objective and latent interface, but differ in how representational capacity is allocated between encoder and decoder. Exact hyperparameters for both settings are reported in \cref{app:titok_hyperparams}.

We report \ac{utmos}~\citep{saeki2022utmos} for clean speech, \ac{dnsmos}~\citep{reddy2022dnsmos} for noisy speech, \ac{dwer}~\citep{wang2021dwer} for linguistic preservation, and speaker similarity (Sim) from a pretrained speaker-verification model. Metric checkpoints and preprocessing details are provided in \cref{app:metric_checkpoints}.

%% file: sections/results.tex
\input{tables/resynthesis}

\section{Results}

\subsection{Resynthesis Quality}
\label{sec:resynthesis}

\cref{tab:speech_resynthesis} reports resynthesis quality on LibriSpeech test-clean, VoiceBank, and Libri1Mix. Across all three benchmarks, we evaluate \textsc{LATTE Large}, the full model with \ac{ola} chunking at a one-second window, and \textsc{LATTE Base}, a smaller asymmetric variant at the standard 50\,Hz token rate.

\paragraph{Clean Speech.}
On LibriSpeech test-clean, \textsc{LATTE Large} achieves an \acs{utmos} of 4.23, the highest among the latent tokenizer variants and comparable to FocalCodec (4.05) and Stable Codec (4.32), the two strongest baselines on this metric.
Speaker similarity (Sim\,=\,97.4) matches FocalCodec, indicating that the latent slot bottleneck preserves speaker identity faithfully despite operating on a non-temporally aligned representation.
The \acs{dwer} of 5.82 is higher than FocalCodec (2.18) and BigCodec (2.55), suggesting a moderate intelligibility cost attributable to the latent-slot compression; the bottleneck must reconstruct fine phonetic detail from global, position-unaligned tokens rather than frame-level codes.

\paragraph{Noisy Speech.}
On VoiceBank, \textsc{LATTE Large} achieves a \acs{dnsmos} of 3.29, second only to Stable Codec (3.33) and above all other baselines, including FocalCodec (3.16).
Its \acs{dwer} of 16.30 is the second-best noisy-speech result, behind FocalCodec (8.08), and substantially better than most higher-bitrate baselines such as EnCodec (28.16) and DAC (63.90).
On Libri1Mix, \textsc{LATTE Large} achieves the highest \ac{dnsmos} (3.03) and ties FocalCodec for best Sim (91.5\,vs.\,91.6), while its \acs{dwer} of 39.07 is again mid-table.

Across all three benchmarks, the gap between \textsc{LATTE Large} and FocalCodec on \ac{dwer} is the most consistent difference: the latent-slot design sacrifices some word-level intelligibility relative to a frame-rate codec at the same bitrate, while matching or exceeding it on perceptual quality and speaker similarity.
This trade-off reflects the core design choice: by discarding temporal frame alignment in favour of a compact set of global tokens, the model gains a representation that is more amenable to token-space manipulation at the cost of slightly looser phonetic reconstruction.

\subsection{Slot-Importance Analysis}
\label{sec:slot_importance}

We next ask whether \textsc{LATTE}'s latent positions behave as interchangeable
containers or develop factor-dependent structure. For each annotated partition,
we compute the TiTok-style slot-importance score from \cref{sec:importance} and
normalize the resulting profile to sum to one across slots. We evaluate speaker
identity, speaker gender, accent, and noise level. Speaker and gender partitions
are computed on both LibriTTS and VCTK; noise partitions are computed on
LibriTTS using white noise and WHAM!~\citep{wichern2019wham} noise across multiple SNRs.

\begin{figure}[t]
    \centering
    \includegraphics[width=.9\linewidth]{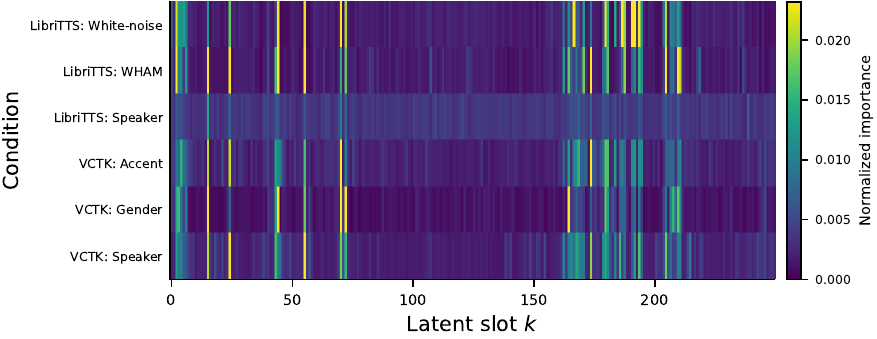}
    \caption{
        Row-normalized slot-importance profiles for different factor partitions.
        Each row is $\ell_1$-normalized across slots. The heatmap shows that
        importance is non-uniform and factor-dependent, with noise-related
        partitions emphasizing similar slot regions across corruption types.
    }
    \label{fig:slot_importance_heatmap}
\end{figure}

\begin{table}[t]
\centering
\caption{
    Structure of slot-importance profiles.
    Left: profile concentration, measured by entropy over the row-normalized
    importance distribution and by the Gini coefficient. Right: similarity
    between selected profile pairs, measured by Spearman correlation over all
    slots and Jaccard overlap among the top-ranked slots. Lower entropy and
    higher Gini indicate stronger concentration. We abbreviate speaker as Spk.,
    LibriTTS as L, and VCTK as V.
}
\label{tab:importance_structure}

\begin{minipage}[t]{0.38\linewidth}
\centering
\begin{tabular}{lcc}
\toprule
\multicolumn{3}{c}{\textbf{Profile concentration}} \\
\midrule
Profile & Ent. $\downarrow$ & Gini $\uparrow$ \\
\midrule
Noise: White      & 4.86 & 0.55 \\
Noise: WHAM!      & 4.78 & 0.57 \\
Speaker: L        & 5.49 & 0.13 \\
Accent: V         & 5.16 & 0.42 \\
Gender: V         & 4.04 & 0.70 \\
Speaker: V        & 5.18 & 0.42 \\
\bottomrule
\end{tabular}
\end{minipage}
\hfill
\begin{minipage}[t]{0.58\linewidth}
\centering
\begin{tabular}{lccc}
\toprule
\multicolumn{4}{c}{\textbf{Profile similarity}} \\
\midrule
Comparison & $\rho$ & Jacc@5 & Jacc@10 \\
\midrule
White vs WHAM!          & 0.735 & 0.111 & 0.177 \\
Spk.: L vs Spk.: V      & 0.308 & 0.429 & 0.429 \\
Spk.: V vs Accent: V    & 0.515 & 0.667 & 0.429 \\
Spk.: V vs Gender: V    & 0.433 & 0.250 & 0.250 \\
\bottomrule
\end{tabular}
\end{minipage}
\end{table}

\Cref{fig:slot_importance_heatmap} shows that slot importance is strongly
non-uniform. Most factors place a considerable share of their mass on a
small subset of latent positions, and the active regions vary by factor. This
rules out a purely exchangeable view of the latent sequence: even though the
tokens are not temporally aligned, their positions acquire distinct functional
roles. At the same time, the profiles are not one-hot or perfectly disjoint,
which already suggests partial specialization rather than hard disentanglement.
\Cref{tab:importance_structure}-left quantifies this pattern with entropy and
the Gini index. Noise-related profiles have similar concentration across white
noise and WHAM! noise. VCTK gender is the most
concentrated profile, whereas LibriTTS speaker identity is substantially more
diffuse, indicating that different attributes occupy the latent sequence at
different sparsity levels.

We then evaluate whether the same factors select stable slots across related
partitions. \Cref{tab:importance_structure}-right reports Spearman correlations and
top-$k$ Jaccard overlaps between selected profile pairs. Noise-related profiles
are the most stable: white-noise and WHAM!-noise partitions have a Spearman
correlation of $0.735$. Thus, the estimator appears to recover slots associated with
degradation level rather than artifacts of a particular corruption process.

Speaker identity, gender, and accent show a more entangled pattern. VCTK speaker
importance overlaps with both accent and gender, which is expected because these
attributes are statistically correlated in the dataset. Speaker-related profiles
also transfer less cleanly across datasets than noise profiles, suggesting that
speaker importance depends more strongly on the population and partition used to
estimate it. We therefore do not interpret the slots as isolated semantic
factors. The evidence supports a weaker and more useful conclusion: the latent
representation has structured, factor-dependent importance profiles, but related
attributes remain partially entangled.

Overall, the latent slots are not interchangeable: different global factors
induce distinct and, for noise, highly stable importance profiles. The profiles
are nevertheless not fully separated across speaker-related attributes. This
motivates the token-space editing experiments below, where we test whether
structured but entangled importance profiles are still useful for targeted
interventions. Cumulative-mass curves and per-factor diagnostics
overlap plots are provided in \cref{app:slot_importance_extra}.

\input{tables/resynthesis_denoising_comparison}
\input{tables/voice_conversion}

\input{sections/discussion}

%% file: tables/resynthesis.tex
\begin{table*}[t]
    \setlength{\tabcolsep}{3pt}
    \caption{Speech resynthesis results on clean and noisy speech benchmarks. Higher UTMOS, DNSMOS, and speaker similarity (Sim) are better; lower dWER is better.}
    \label{tab:speech_resynthesis}
    \centering
    \begin{footnotesize}
    \resizebox{\textwidth}{!}{%
    \begin{tabular}{lc|ccc|ccc|ccc}
    \toprule
    \multirow{2}{*}{\textbf{Codec}} &
    \multirow{2}{*}{\makecell{\textbf{Bitrate} \\ \textbf{(kbps)}} $\downarrow$}
    & \multicolumn{3}{c|}{\textit{Clean – LibriSpeech test-clean}}
    & \multicolumn{3}{c|}{\textit{Noisy – VoiceBank}}
    & \multicolumn{3}{c}{\textit{Noisy – Libri1Mix}} \\
    
    \cmidrule(lr){3-5}
    \cmidrule(lr){6-8}
    \cmidrule(lr){9-11}
    
    & & \textbf{UTMOS} $\uparrow$ & \textbf{dWER} $\downarrow$ & \textbf{Sim} $\uparrow$
      & \textbf{DNSMOS} $\uparrow$ & \textbf{dWER} $\downarrow$ & \textbf{Sim} $\uparrow$
      & \textbf{DNSMOS} $\uparrow$ & \textbf{dWER} $\downarrow$ & \textbf{Sim} $\uparrow$ \\
    
    \midrule
    Reference       & ---  & 4.09 & 0.00 & 100.0 & 3.56 & 0.00 & 100.0 & 3.73 & 0.00 & 100.0 \\
    
    EnCodec         & 1.50 & 1.58 & 8.08 & 93.8  & 2.76 & 28.16 & 87.7 & 2.40 & 55.17 & 86.3 \\
    
    DAC             & 1.00 & 1.29 & 20.04 & 89.2 & 2.72 & 63.90 & 79.8 & 2.40 & 90.92 & 76.6 \\
    
    WavLM6-KM       & 0.45 & 3.75 & 6.20 & 90.0 & 3.06 & 20.67 & 82.9 & 2.87 & \underline{36.60} & 85.9 \\
    
    SpeechTokenizer & 1.00 & 2.28 & 5.14 & 91.6 & 2.74 & 34.51 & 82.2 & 2.58 & 57.26 & 82.8 \\
    
    SemantiCodec    & 0.65 & 2.91 & 8.97 & 96.0 & 3.13 & 31.46 & 90.6 & 2.67 & 51.18 & 89.9 \\
    
    Mimi            & 0.69 & 3.29 & 5.73 & 96.0 & 3.01 & 28.00 & 87.8 & 2.65 & 49.14 & 89.4 \\
    
    WavTokenizer    & 0.48 & 3.78 & 11.55 & 95.4 & 3.09 & 42.12 & 89.8 & 2.53 & 70.10 & 86.3 \\
    
    BigCodec        & 1.04 & 4.11 & \underline{2.55} & \textbf{98.5}
                    & 3.19 & 20.67 & \textbf{92.3}
                    & 2.75 & 53.26 & 88.3 \\
    
    Stable Codec    & 0.70 & \textbf{4.32} & 4.97 & 94.7
                    & \textbf{3.33} & 20.32 & 88.8
                    & 2.91 & 43.52 & 90.0 \\
    
    FocalCodec
                    & 0.65 & 4.05 & \textbf{2.18} & \underline{97.4}
                    & 3.16 & \textbf{8.08} & 91.3
                    & {2.93} & \textbf{27.89} & \underline{91.6} \\
    
    \midrule
    
    \textbf{\textsc{\textsc{LATTE Large}}}
                    & 0.65 & \underline{4.23} & 5.82 & \underline{97.4}
                    & \underline{3.29} & \underline{16.30} & \underline{91.6}
                    & \textbf{3.03} & 39.07 & {91.5} \\
    
    \textbf{\textsc{\textsc{LATTE Base}}}
                    & 0.65 & 4.20 & 7.07 & 96.9
                    & 3.26 & 19.51 & 91.0
                    & \underline{2.96} & 41.44 & \textbf{91.7} \\
    
    \bottomrule
    \end{tabular}
    }
    \end{footnotesize}
    \vskip -10pt
\end{table*}

%% file: tables/resynthesis_denoising_comparison.tex
\begin{table*}[t!]
    \setlength{\tabcolsep}{3pt}
    \caption{Noisy-set resynthesis compared with mass-based noise-slot replacement. Each corpus reports \textsc{LATTE Large} and mass-based slots at cumulative thresholds $\gamma\in\{0.02,\,0.10\}$. Noise type appears as two four-column groups on each row (\textbf{WHAM} vs \textbf{white noise}), with DNSMOS, \ac{dwer}, and Sim (higher DNSMOS and Sim are better; lower \ac{dwer} is better). The resynthesis baseline is noise-family agnostic, so DNSMOS\,/\,\ac{dwer}\,/\,Sim are repeated under both groups.}
    \label{tab:resynthesis_denoising_comparison}
    \centering
    \vspace{5pt}
    \begin{footnotesize}
    \begin{tabular}{@{}ll ccc ccc@{}}
    \toprule
    & & \multicolumn{3}{c}{\textbf{WHAM}} & \multicolumn{3}{c}{\textbf{White noise}} \\
    \cmidrule(lr){3-5} \cmidrule(lr){6-8}
    \textbf{Corpus} & \textbf{Setting} & \textbf{DNSMOS} $\uparrow$ & \textbf{dWER} $\downarrow$ & \textbf{Sim} $\uparrow$ & \textbf{DNSMOS} $\uparrow$ & \textbf{dWER} $\downarrow$ & \textbf{Sim} $\uparrow$ \\
    \midrule
    \multirow{3}{*}[-2pt]{\textit{VoiceBank}} 
    & \textsc{LATTE (Large)} & 3.29 & 16.30 & \textbf{91.6} & 3.29 & 16.30 & \textbf{91.6} \\
    & $\gamma{=}0.02$ & \underline{3.50} & \underline{11.23} & 89.8 & \underline{3.59} & \textbf{11.52} & \underline{90.8} \\
    & $\gamma{=}0.10$ & \textbf{3.58} & \textbf{9.54} & \underline{90.4} & \textbf{3.60} & \underline{11.61} & 90.7 \\
    \midrule
    \multirow{3}{*}[-2pt]{\textit{Libri1Mix}} 
    & {\textsc{LATTE (Large)}} & 3.03 & 39.07 & \textbf{91.5} & 3.03 & 39.07 & \textbf{91.5} \\
    & $\gamma{=}0.02$ & \underline{3.14} & \textbf{30.77} & \underline{89.1} & \underline{3.40} & \underline{36.07} & 87.9 \\
    & $\gamma{=}0.10$ & \textbf{3.36} & \underline{35.13} & 86.3 & \textbf{3.41} & \textbf{31.26} & \underline{88.0} \\
    \bottomrule
\end{tabular}
        \end{footnotesize}
\end{table*}

%% file: tables/voice_conversion.tex
\begin{table*}[t]
\centering
\setlength{\tabcolsep}{4pt}
\caption{
    One-shot parallel voice conversion on VCTK~\citep{veaux2017cstr}.
    We report reference systems and codec baselines in the first block, followed by
    \textsc{LATTE Large} token-swap results on
    \texttt{vctk\_parallel\_test\_mic1}. For token swapping, $\gamma$ denotes the
    cumulative-mass threshold and $k$ the number of swapped tokens.
    Importance-guided token selection improves target-speaker similarity over
    random and least-important controls at matched edit budgets.
}
\label{tab:voice_conversion}
\vspace{0.1cm}
 \resizebox{0.6\textwidth}{!}{%
\begin{tabular}{lcccccc}
\toprule
\textbf{Method} &
\textbf{$\gamma$} &
\textbf{$k$} &
\makecell{\textbf{Bitrate} \\ \textbf{(kbps)} $\downarrow$} &
\textbf{UTMOS} $\uparrow$ &
\textbf{dWER} $\downarrow$ &
\textbf{Sim} $\uparrow$ \\
\midrule
\multicolumn{7}{l}{\textit{Reference and codec baselines}} \\
\midrule
Reference        & --- & --- & ---  & 4.09 & 0.00   & 100.0 \\
EnCodec          & --- & --- & 1.50 & 1.24 & 86.52  & 72.2 \\
DAC              & --- & --- & 1.00 & 1.25 & 104.00 & 67.2 \\
WavLM6-KM         & --- & --- & \textbf{0.45} & 2.90 & 26.68  & \textbf{92.4} \\
SpeechTokenizer  & --- & --- & 1.00 & 1.49 & \underline{20.32} & 81.2 \\
SemantiCodec     & --- & --- & 0.65 & 2.02 & 106.00 & 72.8 \\
Mimi             & --- & --- & 0.69 & 2.40 & 110.00 & {89.7} \\
WavTokenizer     & --- & --- & \underline{0.48} & 3.13 & 43.15  & 73.4 \\
BigCodec         & --- & --- & 1.04 & 1.31 & 99.96  & 68.9 \\
Stable Codec     & --- & --- & 0.70 & 3.76 & 27.63 & 71.1 \\
FocalCodec       & --- & --- & 0.65 & 3.38 & {21.27} & \underline{92.2} \\
\midrule
\multicolumn{7}{l}{\textit{Importance-guided token swap}} \\
\midrule
\textbf{\textsc{LATTE (Large)}} & 0.05 & 3 & 0.65 & 4.15 & \textbf{11.24} & 89.5 \\
\textbf{\textsc{LATTE (Large)} }& 0.07 & 4 & 0.65 & 4.14 & 12.53 & 89.6 \\
\textbf{\textsc{LATTE (Large)}} & 0.10 & 5 & 0.65 & \textbf{4.16} & 12.32 & 90.0 \\
\midrule
\multicolumn{7}{l}{\textit{Matched-budget editing controls}} \\
\midrule
Random control      & 0.10 & 5  & --- & 4.08 & 16.26 & 67.8 \\
Least control       & 0.10 & 5  & --- & \underline{4.10} & 12.03 & 68.1 \\
\bottomrule
\end{tabular}
}
\vspace{-20pt}
\end{table*}

%% file: sections/discussion.tex
\subsection{Token-Space Editing}
\label{sec:token_space_editing}

We evaluate denoising and parallel voice conversion by replacing
importance-ranked latent slots with tokens from a reference utterance
(\cref{tab:resynthesis_denoising_comparison,tab:voice_conversion}). These
experiments are controlled interventions on the tokenizer, not fully trained
conversion or enhancement systems.

\paragraph{Denoising.}
Mass-based noise-slot replacement \cref{eq:swap1,eq:slot2} improves perceptual quality on both noisy
benchmarks. On VoiceBank, \ac{dwer} drops from 16.30 under no-swap
resynthesis to roughly 9.5--11.6 for the reported settings, while \ac{mos}
also increases. On Libri1Mix, \ac{mos} increases and \ac{dwer} generally falls
into the low-to-mid 30s, a smaller but consistent gain under harder mixture
conditions. Speaker similarity decreases slightly because clean references are  generally unmatched in text and speaker.

\paragraph{Voice Conversion.}
Mass-based speaker-slot swaps reach high \ac{utmos} while keeping
parallel-VCTK \ac{dwer} in the single-digit-to-low-teens range. Increasing the
number of replaced slots raises target-speaker similarity but can increase
\ac{dwer}, revealing the expected edit-strength trade-off. We compare against two matched-budget controls: a random control, which swaps the same number of token slots chosen uniformly at random, and a least-important control, which swaps the same number of slots with the lowest importance scores. Both controls yield substantially lower target-speaker similarity, suggesting that the importance scores identify functionally relevant slots rather than arbitrary perturbation directions. Random and
least-important controls at matched budgets produce much lower target
similarity, supporting the claim that the importance scores identify
functionally relevant slots rather than arbitrary perturbation directions. Additional protocol details and interpretation guidelines are
provided in \cref{app:editing_protocols}.

%% file: sections/appendix_editing_protocols.tex
\section{Token-Swap Protocol Details}
\label{app:editing_protocols}

\paragraph{Cumulative-Importance Slot Replacement as an Evaluation Protocol.}
Using cumulative-importance replacement turns slot importance into a testable
intervention: if a factor is concentrated in a subset of slots, replacing the
smallest set of top-ranked slots that accounts for a fraction $\gamma$ of the
total importance should move task-specific metrics while preserving non-target
attributes. Varying $\gamma$ provides a controllable edit strength. Small values
of $\gamma$ probe highly concentrated, high-confidence slots, while larger values
test whether useful signal is distributed across a broader set of slots. In this
sense, cumulative-importance replacement functions as a causal probe of
representational structure rather than a recipe for optimal downstream quality.

\paragraph{Speaker Editing with Parallel VCTK References.}
For speaker editing, the reference is a different speaker reading the same
 This parallel condition reduces lexical confounds and makes Sim and
\ac{dwer} changes easier to attribute to speaker-token transfer instead of
content mismatch. The setup is intentionally favourable for analyzing speaker
transfer under controlled conditions; more realistic non-parallel conversion
remains an important extension.

\paragraph{Denoising.}
For denoising, the target utterance is clean and randomly selected from the clean subset; it is therefore typically unmatched in both transcript and speaker. This prevents trivial identity-copy behavior and avoids relying on paired clean targets. Any gains from slot replacement therefore indicate that the identified noise-associated slots can be manipulated robustly through token swapping.

\paragraph{Importance-Guided vs Control Slot Selections.}
Random slots and least-important slots are the critical
controls. Random and least-important controls test whether gains come from
targeted factor localization rather than generic perturbation. If importance-guided edits dominate these controls at matched $\gamma$, this supports the claim that
slot-importance profiles identify functionally relevant positions.

\paragraph{No-Swap Source Reconstruction as a Reference.}
The no-swap setting reconstructs the source utterance without slot replacement.
It provides a reconstruction baseline, helping distinguish the effect of swapping
from artifacts already introduced by the tokenizer and decoder.

\paragraph{Asymmetric Encoder--Decoder Capacity.}
Our architecture uses a lighter encoder than decoder. This choice is motivated
by generative workloads, where encoding is often performed offline or at large
scale for preprocessing, augmentation, and conditioning extraction, while
decoding quality directly controls perceptual fidelity. A stronger decoder
better handles the one-to-many inverse mapping from compressed latents to
waveform detail, whereas a leaner encoder reduces front-end compute. This
asymmetry follows the same motivation as recent generative-first neural audio
autoencoders, where encoder-side efficiency and decoder capacity are central to
tokenization throughput and downstream usability \citep{11463285}.

%% file: sections/appendix_ola.tex
\section{Overlap-Add Inference}
\label{app:ola}

Long utterances may exceed the fixed temporal context used to train
\textsc{LATTE}. At inference time, we therefore process long recordings with
feature-domain overlap-add (OLA). The waveform is first encoded into
FocalCodec features, overlapping feature chunks are passed through
\textsc{LATTE}, and the resulting reconstructed feature chunks are merged
before a single vocoder pass.

Let $x[n]$ be a mono waveform sampled at \SI{16}{\kilo\hertz}. The FocalCodec
encoder maps it to a feature sequence
$\mathbf{X}\in\mathbb{R}^{T\times D}$ at \SI{50}{\hertz}, corresponding to one
feature frame every \SI{20}{\milli\second}. We use feature chunks matching the
training chunk duration. If the training chunk contains $c$ waveform samples,
the corresponding feature length is
\[
    K = \mathrm{round}\!\left(\frac{c}{f_s}\cdot 50\right),
\]
where $f_s=\SI{16000}{\hertz}$.

For long utterances, we slide this window over the feature sequence with a
fixed overlap of \SI{1}{\second}, i.e., $50$ feature frames. Thus the hop size is
\[
    h = K - 50.
\]
The feature sequence is right-padded when needed so that the final window fits
the regular sliding-window grid. For each window $i$, we extract a feature chunk
$\mathbf{X}^{(i)}\in\mathbb{R}^{K\times D}$, run it through
\textsc{LATTE} quantization and decoding, and obtain a reconstructed feature
chunk $\widetilde{\mathbf{X}}^{(i)}$.

Overlapping chunks are merged using a non-periodic Hann window
$\mathbf{w}\in\mathbb{R}^{K}$:
\[
    \boldsymbol{\Omega} \leftarrow
    \boldsymbol{\Omega} + \widetilde{\mathbf{X}}^{(i)} \odot \mathbf{w},
    \qquad
    \omega \leftarrow \omega + \mathbf{w},
\]
where $\mathbf{w}$ is broadcast over the feature dimension and $\omega$ stores
the accumulated window envelope. The stitched feature sequence is then
\[
    \widetilde{\mathbf{X}}
    =
    \frac{\boldsymbol{\Omega}}{\mathrm{clamp}(\omega, 10^{-8})}.
\]
After fusion, we crop the sequence back to the original feature length $T$ and
decode the full stitched feature trajectory with the FocalCodec vocoder. The
decoded waveform is finally cropped to the original waveform length before
evaluation.

For utterances shorter than the training chunk, we right-pad the waveform,
process it with a single \textsc{LATTE} forward pass, and crop the decoded audio
back to the original duration. This avoids applying temporal fusion when only
one chunk is needed.

%% file: sections/appendix_titok_hyperparams.tex
\section{TiTok Training Hyperparameters}
\label{app:titok_hyperparams}

This appendix reports the hyperparameters used for the TiTok-style \textsc{LATTE Base} and \textsc{LATTE Large}
tokenizer variants. Unless otherwise stated, both variants use the same training, optimization, data, and runtime settings. The only architectural difference between \textsc{LATTE Base} and \textsc{LATTE Large} is the width preset used for the encoder and decoder compressor stacks. All \textsc{LATTE} models used in this paper use a 50 token per second rate.

\subsection{Model Variants}
\label{app:titok_variants}

Table~\ref{tab:titok_variants} summarizes the architectural presets used for the
two variants. We use the shorthand BL to denote a base encoder with a large
decoder, and \textsc{LATTE Large} to denote a large encoder with an XL decoder.

\begin{table}[h]
\centering
\caption{TiTok-style model variants used in our experiments. Width presets are
reported as \texttt{hidden\_dims} for the corresponding compressor or
decompressor stack.}
\label{tab:titok_variants}
\begin{tabular}{lcc}
\toprule
Variant & Encoder width preset & Decoder width preset \\
\midrule
BL  & \texttt{base}: $(1024, 512, 256)$  & \texttt{large}: $(2048, 1024, 512)$ \\
\textsc{LATTE Large} & \texttt{large}: $(2048, 1024, 512)$ & \texttt{xl}: $(4096, 2048, 1024)$ \\
\bottomrule
\end{tabular}
\end{table}

All compressor and decompressor temporal factors are set to $1$, so the TiTok
compressor stack does not perform additional time-axis scaling.

\subsection{Shared Model Hyperparameters}
\label{app:titok_model}

Table~\ref{tab:titok_model_hparams} reports the model hyperparameters shared by
the \textsc{LATTE Base} and \textsc{LATTE Large} variants.

\begin{table}[h]
\centering
\caption{Shared model hyperparameters for the \textsc{LATTE Base} and \textsc{LATTE (Large)} variants.}
\label{tab:titok_model_hparams}
\begin{tabular}{ll}
\toprule
Hyperparameter & Value \\
\midrule
Input feature dimension & $1024$ \\
Quantizer & Binary spherical quantizer (BSQ) \\
Code dimension & $13$ \\
Chunk duration & \SI{5}{s} \\
Maximum sequence length (Positional Embedding) & $10{,}000$ \\
\bottomrule
\end{tabular}
\end{table}

\subsection{Shared Optimization Hyperparameters}
\label{app:titok_optim}

Table~\ref{tab:titok_optim_hparams} reports the optimization hyperparameters
used for both variants.

\begin{table}[h]
\centering
\caption{Shared optimization hyperparameters for LATTE training.}
\label{tab:titok_optim_hparams}
\begin{tabular}{ll}
\toprule
Hyperparameter & Value \\
\midrule
Optimizer & AdamW \\
Learning rate & $5 \times 10^{-4}$ \\
Adam betas & $(0.9, 0.98)$ \\
Weight decay & $0.01$ \\
Gradient clipping & Max norm $5.0$ \\
BSQ entropy auxiliary loss weight & $0.1$ \\
BSQ inverse temperature & $100.0$ \\
Diversity weight & $1.0$ \\
\bottomrule
\end{tabular}
\end{table}

The learning rate is scheduled with \texttt{ReduceLROnPlateau}. The scheduler is
configured with \texttt{mode=min}, factor $0.9$, patience $0$, improvement
threshold $0.0025$, and minimum learning rate $10^{-6}$.

\subsection{Data and Dataloader Settings}
\label{app:titok_data}

Table~\ref{tab:titok_data_hparams} reports the data and dataloader settings.

\begin{table}[]
\centering
\caption{Data and dataloader settings used for LATTE training.}
\label{tab:titok_data_hparams}
\begin{tabular}{ll}
\toprule
Setting & Value \\
\midrule
Training splits & \texttt{train-clean-100}, \texttt{train-clean-360}, \texttt{train-other-500} \\
Validation split & \texttt{dev-clean} \\
Dataset & LibriTTS \\
Audio sampling rate & \SI{16}{kHz} \\
Chunk size & $80{,}000$ samples \\
Chunk duration & \SI{5}{s} \\
Number of workers & $4$ \\
Batch size & $4$ \\
\bottomrule
\end{tabular}
\end{table}

The chunk size corresponds to $\SI{5}{s} \times \SI{16000}{Hz}$.

\subsection{Computational Resources}
The model is trained on a node with 4x H100 NVIDIA GPUs (80 GB).